\documentclass[11pt]{article}
\usepackage{times}
\usepackage{geometry}
\usepackage[utf8]{inputenc}
\usepackage{enumitem,amssymb}
\usepackage{ragged2e}
\newlist{thematic}{itemize}{8}
\setlist[thematic]{label=$\square$}
\usepackage{pifont}
\usepackage{fancyhdr}
\usepackage[USenglish]{babel}
\usepackage[utf8]{inputenc}
\usepackage[T1]{fontenc}
\usepackage{epsfig}
\usepackage{wrapfig}
\usepackage{color}
\usepackage{framed}

\usepackage[vflt]{floatflt}
\usepackage{longtable}
\usepackage{caption}
\usepackage{jheppub}   
\usepackage[dvipsnames,svgnames,x11names]{xcolor}
\usepackage[all]{hypcap} 
\usepackage{amssymb,amsmath}
\usepackage{longtable}
\usepackage{amssymb}
\usepackage{amsmath, xspace}
\usepackage{graphics,graphicx} 
\usepackage{rotating}
\usepackage{color}
\usepackage{xcolor}
\usepackage{multirow}
\usepackage{subfigure}
\usepackage{sectsty}
\usepackage[outercaption]{sidecap}
\sidecaptionvpos{figure}{c}

\definecolor{DarkGreen}{rgb}{0.0, 0.3, 0.0}
\definecolor{purple}{rgb}{0.5, 0.0, 0.5}
\definecolor{red}{rgb}{1, 0.0, 0.0}
\definecolor{green}{rgb}{0, 1.0, 0.0}

\newcommand{\Moon}{$M_{Moon}$}                          

\newcommand*\aap{A\&A}

\newcommand*\ao{Appl.~Opt.}

\newcommand*\apjl{ApJ}

\newcommand*\apjs{ApJS}

\def\ao{{\it Appl.~Opt.}}           

\def\3he{$^3{\rm He}$}
\def\arcdeg{\hbox{$^\circ$}}

\def\lsim{\mathrel{\lower2.5pt\vbox{\lineskip=0pt\baselineskip=0pt
           \hbox{$<$}\hbox{$\sim$}}}}

\def\gsim{\mathrel{\lower2.5pt\vbox{\lineskip=0pt\baselineskip=0pt
           \hbox{$>$}\hbox{$\sim$}}}}

\usepackage[citestyle=numeric-comp,bibstyle=science,hyperref=true,uniquename=mininit,uniquelist=false,natbib=true,defernumbers=true,eprint=false,sorting=none,maxcitenames=1,maxbibnames=2,]{biblatex}
\makeatletter
\newcommand*{\blx@fnpct@movefor}{}
\newcommand*{\DeclareFootnoteMovePunct}{%
  \@ifstar
    {\let\blx@fnpct@movefor\@empty}
    {}
  \blx@def@fnpct@movefor}
\newcommand*{\blx@def@fnpct@movefor}{%
  \def\do##1{\blx@thecheckpunct{\listadd{\blx@fnpct@movefor}}##1}%
  \docsvlist}
\DeclareFootnoteMovePunct{.,{,}}
\newlength{\blx@fnpct@movelength}
\newcommand*{\blx@fnpct@footnotemover}[1]{%
  #1%
  \ifinlist{#1}{\blx@fnpct@movefor}
    {\settowidth{\blx@fnpct@movelength}{#1}%
     \hspace{-1.\blx@fnpct@movelength}}
    {}%
}
\protected\csedef{blx@acitei@superscript}#1#2#3#4#5{%
  \begingroup
  \blx@citeinit
    \noexpand\blx@fnpct@footnotemover{#5}%
    \blxcitecmd{supercite#1}{#2}{#3}{#4}{}%
  \endgroup}%
\protected\csedef{blx@macitei@superscript}#1#2#3{%
  \noexpand\blx@fnpct@footnotemover{#3}%
  #1{#2}%
  \endgroup}
\makeatother

\DeclareFieldFormat[article]{volume}{#1}
\DeclareFieldFormat[article]{journaltitle}{#1}

\AtEveryBibitem{\clearfield{doi}}
\AtEveryCitekey{\clearfield{doi}}

\AtEveryBibitem{\clearfield{url}}
\AtEveryCitekey{\clearfield{url}}

\AtEveryBibitem{\clearfield{month}}
\AtEveryCitekey{\clearfield{month}}

\DeclareBibliographyDriver{arxiv}{%
    \printnames{author}%
    \newunit\newblock%
    \printfield{eprint}%
    \setunit{\space}%
    \printfield{year}
}

\DeclareFieldFormat[arxiv]{title}{{#1}}
\DeclareFieldFormat[arxiv]{year}{({#1})}
\DeclareFieldFormat[arxiv]{eprint}{arXiv:{#1}}

\renewbibmacro*{name:andothers}{%
  \ifboolexpr
    {
      test {\ifnumequal{\value{listcount}}{\value{liststop}}}
      and
      test \ifmorenames
    }
    {
      \ifnumgreater{\value{liststop}}{1}
        {\finalandcomma}
        {}%
      \andothersdelim
      \bibstring{andothers}%
    }
    {}%
}

\DeclareFieldFormat{labelnumberwidth}{[#1]}

\defbibenvironment{bibliography}
  {}
  {}
  {\addspace
   \printtext[labelnumberwidth]{%
     \printfield{prefixnumber}%
     \printfield{labelnumber}}%
   \addhighpenspace}
   
\renewbibmacro*{finentry}{\finentry\quad}

\begin{document}
{\raggedright
\huge
Sub-millimeter wavelength protostellar accretion rate monitoring with AtLAST
\linebreak
\bigskip
\normalsize



\bigskip

\textbf{Authors:}\\
{\bf Thomas Stanke} (tstanke@mpe.mpg.de, Max-Planck-Institute for Extraterrestrial Physics, Germany)\\
{\bf Verena Wolf} (Th\"uringer Landessternwarte Tautenburg, Germany)\\
{\bf Bringfried Stecklum} (Th\"uringer Landessternwarte Tautenburg, Germany)\\
{\bf Doug Johnstone} (NRC Herzberg Astronomy and Astrophysics Research Centre \& University of Victoria, Canada);
{\bf Jochen Eisl\"offel} (Th\"uringer Landessternwarte Tautenburg, Germany);
{\bf Greg Herczeg} (Kavli Institute for Astronomy and Astrophysics, Peking University, PR China);
{\bf S.\ Tom Megeath} (University of Toledo, USA);
{\bf Karri Koljonen} (Finnish Centre for Astronomy with ESO, Finland \& NTNU, Norway)
\linebreak

\textbf{Science Keywords:} 
stars: formation - stars: protostars - stars: variables\\
}

 \captionsetup{labelformat=empty}
\begin{figure}[h]
   \centering
\includegraphics[width=.9\textwidth]{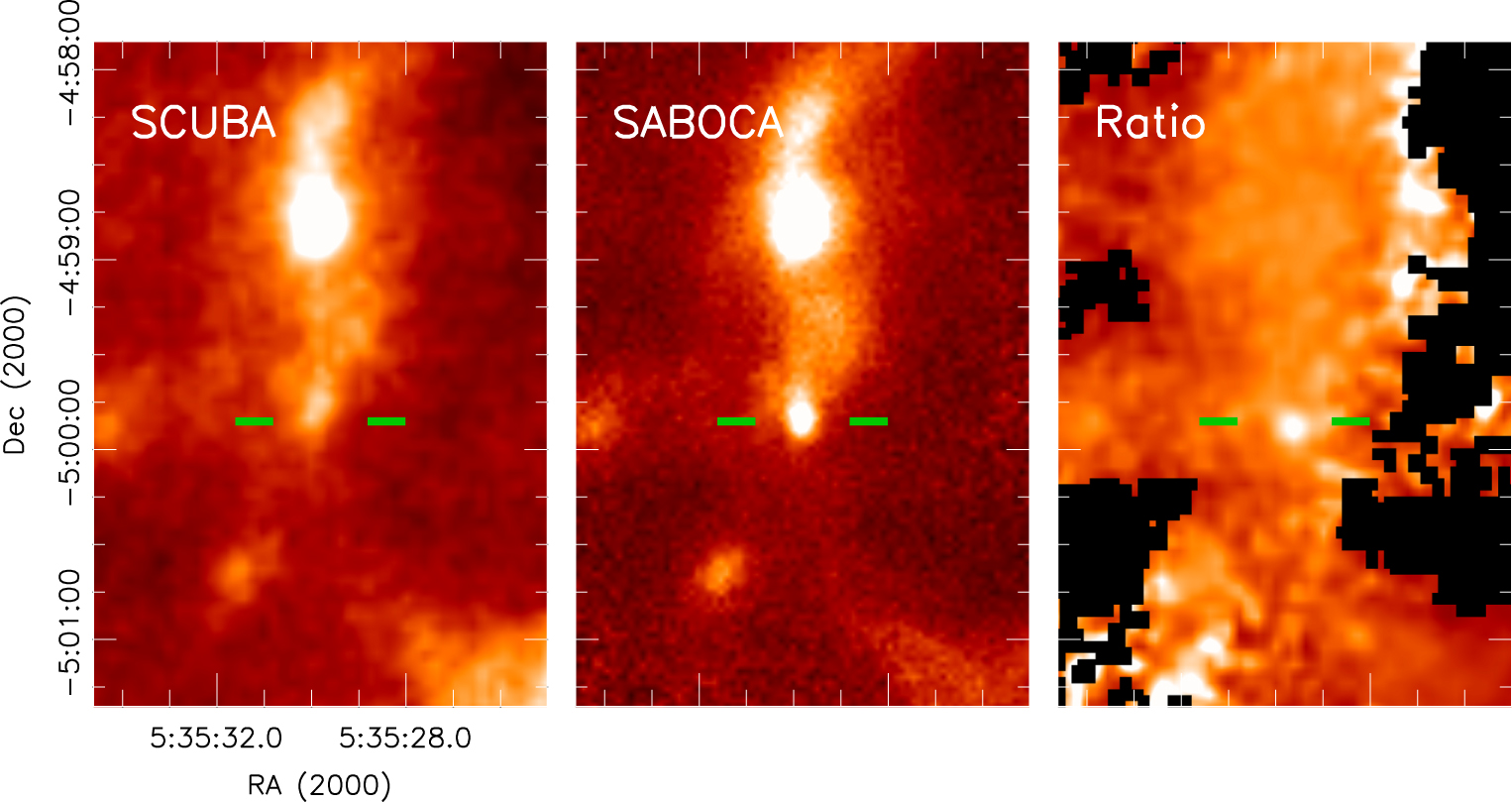}
\caption{The first detected accretion burst in a Class~0 protostar
  (HOPS~383) as seen at sub-millimeter wavelengths (taken from Safron et
  al.\ 2015 \cite{Safron2015}). The outburst started some time between October
  2004 and October 2006. Left: JCMT/SCUBA 450~$\mu$m image taken in 1998, middle:
  APEX/SABOCA 350~$\mu$m image taken in 2011, right: ratio of post- to pre-burst
  image. Only a substantial rise in sub-millimeter flux due to the accretion
  outburst (detected at mid-IR wavelengths) can explain the ratio of the
  post-burst 350~$\mu$m to pre-burst 450~$\mu$m flux seen for HOPS~383. Later
monitoring (2016-2020, \cite{Lee2021}) showed the sub-millimeter flux decreasing by a few percent per year.}
\end{figure}
\vspace{-15mm}

\setcounter{figure}{0}
\captionsetup{labelformat=default}


\pagebreak


\section*{Abstract}

\vspace*{-3mm}
How a star forms is a fundamental question in astrophysics. In the earliest
stages of protostellar evolution high extinction prevents a direct study
of the accretion processes and their temporal evolution. Monitoring the
variations of the accretion luminosity in a large protostar sample over decades
is needed to reveal how protostars accrete -- in major bursts or in a
quasi-steady fashion. We here argue that a large ground based
sub-millimeter single-dish facility with a wide FoV is required to fulfill
this task.

\section{Scientific context and motivation}

\vspace*{-2mm}
Stars form, when the densest regions of molecular clouds -- cloud cores --
get unstable and collapse under their own gravity. At the center of the
collapsing core a hydrostatic central objects forms, surrounded by an accretion
disk, a larger envelope, the collapsing core, and the parent molecular cloud.
Particularly during the earliest -- Class~0 -- phase the central object and its
immediate surroundings are heavily extincted by the surrounding dusty material.
During the subsequent Class~I phase the envelope gradually dissipates and the
central regions become observable at infrared and eventually optical
wavelengths.

However, it is during the short (1-2$\times$10$^5$~yr \cite{Dunham2014}), highly obscured Class~0
phase that the central object gains around 50\% of its final mass!
Due to the high obscuration, traditional UV/optical/near-IR proxys of accretion
cannot be applied. Consequently, the
mechanism of accretion (e.g., magnetospheric vs.\ boundary layer) as well as
actual accretion rates onto protostars are hardly constrained.

As an alternative, the total luminosity of a protostar can be used to constrain
the mass accretion rate, noticing that the total luminosity is the sum of the
intrinsic luminosity of the central hydrostatic object and the energy released
as material accretes onto it. Comparing models (with constant or smoothly
varying accretion rates) with observations shows a tension (the long-standing
``protostellar luminosity problem''), with observed luminosities being too low
compared to predictions \cite{Kenyon1990}. Time variable -- episodic -- accretion, known to
occur in more evolved stages and recently observed in the earliest protostellar
stages, can be a (long suggested) solution to the problem: protostars would
spend most of the time in a low-accretion, low-luminosity state, and only
occasionally undergo phases of high accretion and luminosity.


\vspace{-3mm}
\section{Science case}

\vspace*{-2mm}
Due to the embedded nature of protostars measuring the rate and
temporal evolution of accretion during the earliest phases requires observations
at far-IR to (sub)millimeter wavelengths. The JCMT Transient Survey \cite{Herczeg2017, Lee2021, Mairs2024} offers an example of what can be done today, and the severe
limitations of present generation small telescopes. We here argue that a large
(sub)millimeter single dish telescope providing good spatial resolution along
with a large field-of-view and being available on timescales of decades
(as envisaged by the AtLAST project) is indispensible
in deciphering the accretion process during the early star formation process.

Episodic accretion events are well documented in the later, disk-dominated
stages of star formation. FU~Ori type outburst may last decades to (possibly)
centuries (e.g., \cite{Contreraspena2025}) -- the eponymous object FU~Orionis went into outburst in 1936 and
remained in an elevated state since then \cite{Clarke2005}. FUOr accretion rates are 2 to 3
orders of magnitude higher than in quiescence, causing the objects to brighten
by several magnitudes in the optical. Due to their high accretion rate they may
contribute a very significant fraction of accretion, although their occurence
frequency is not very well constrained (once every few 10$^3$ to few 10$^4$~yr).
EXOr outbursts (named after EX~Lupi) are shorter (months to years) and have
lower amplitudes, but are more frequent (indeed, recurring EXOr outbursts have
been observed, e.g., \cite{Wang2023}).

For the youngest low-mass protostars observational evidence so far
is sparse. Class~0 objects have been directly observed to go through
outbursts only a little more than a decade now \cite{Safron2015},
while episodic protostellar outflow activity
hinted at episodic accretion already before.
Systematic monitoring so far is mostly limited to nearby star forming
regions at mid-IR (e.g., \cite{Zakri2022}) and sub-millimeter wavelengths
(JCMT: \cite{Herczeg2017, Lee2021, Mairs2024, Park2024}), typically
covering a few hundred protostars. While about $1/3$ of the JCMT targets
show some variability, only a few candidates for significant accretion bursts
are seen. While statistics is even more limited than for
more evolved objects, there is evidence that outbursts might
be more frequent early on (e.g., \cite{Fischer2012, Fischer2019, Zakri2022}).
Similarly, strong infall is predicted to trigger more frequent outbursts
at early stages (e.g., \cite{Vorobyov2005}).

In the high mass regime observational evidence is equally sparse, with a good
handfull of accretion bursts observed \cite{Wolf2024, Elbakyan2024, Yang2025}.
For sources with high enough luminosity, particularly the IR
radiation pumped Class~II methanol masers are a very powerful indicator
of accretion variability (e.g., \cite{Stecklum2021}), but maser monitoring
might only be applicable to luminous high-mass sources and fail in the
intermediate- to low mass regime.

{\it   Overall, a tracer of accretion variability for the full range of masses, to be
monitored over decade timescales in substantial protostar samples ---} nearby
star foming regions for low masses, Galactic plane for intermediate and high
masses --- {\it is required
to measure the frequency, duration, and amplitude of accretion bursts as a
function of protostellar mass and evolutionary stage to investigate their
importance relative to (quasi)steady accretion.}

\noindent {\bf The need for far-IR to sub-millimeter observations:} Major
accretion episodes such as EXOr or FUOr events imply elevated temperatures in
the inner disk region. In the deeply embedded stages the radiation emitted by
the hot inner disk is not directly visible. While the outburst might be
observable in scattered light at short wavelengths (e.g., with SPHEREx), the amplitude of the flux
variation is highly dependent on geometry (and foreground extinction). However,
the protostellar envelope will intercept and re-emit the radiation at longer
wavelengths where the surrounding material is increasingly optically thin
\cite{Johnstone2013}.
\vspace*{-2mm}

\begin{figure*}[h]
  \centering
  
\begin{minipage}{0.28\textwidth}
Fischer ea.\ 2024 \cite{Fischer2024} show that the mid-IR
(Spitzer IRAC, WISE, SPHEREx)
is good for detecting outbursts,
but due to the dependence on the scattering and extinction geometry the
correlation between flux and accretion variation shows a large scatter (Fig.\ 1).
The far-IR regime (25-235\,$\mu$m) is shown to most directly reflect the
increase in accretion rate as an increase in luminosity, followed by the
sub-millimeter regime, where the rise in flux for a given rise in accretion
rate decreases towards longer wavelengths.
\end{minipage}
\hfill
\begin{minipage}{0.7\textwidth}
  \centering
  \includegraphics[width=0.9\linewidth]{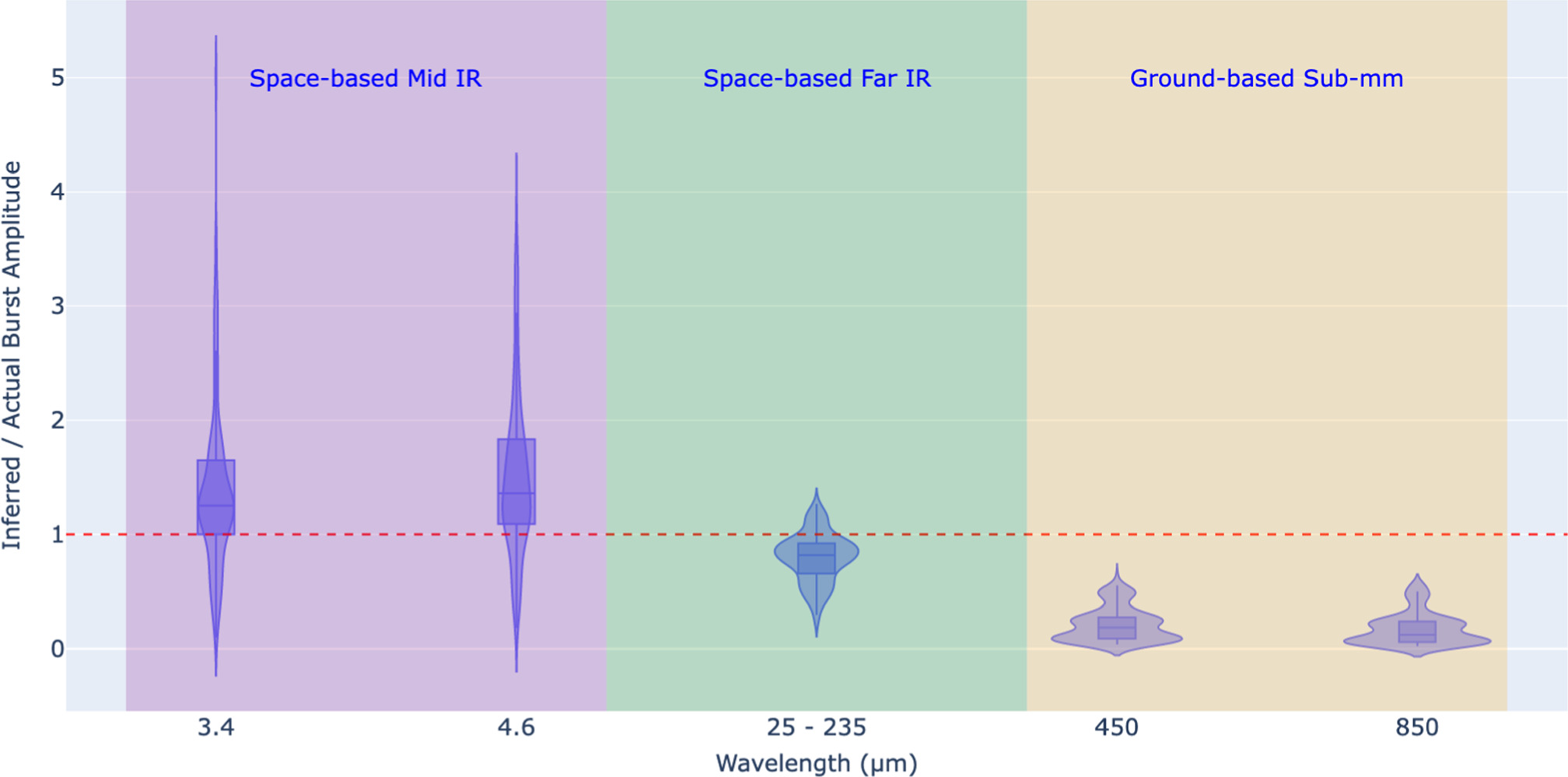}
  \caption{\small Fischer ea.\ 2024 \cite{Fischer2024} Fig.\ 3:
    ratio of
    observed flux variation to actual burst amplitude
    at various wavelengths. The horizontal lines in the violin plots mark the
    first quartile (bottom), median (central), and third quartile (top) of
    each distribution. Space-based (PRIMA) far-IR observations trace the bursts
    most directly with small scatter, followed by sub-millimeter
    observations. The mid-IR range is sensitive to bursts, but shows a large
    scatter of observed vs.\ actual burst amplitude due to a strong dependency
  on source and viewing geometry.}
\end{minipage}
\end{figure*}

\vspace*{-3mm}
\noindent{\bf Why *sub*millimeter monitoring is needed:} the Earth's atmosphere
is opaque at far-IR wavelengths, requiring airborne or space-based
observatories. Currently no far-IR observatory is operational, while PRIMA
(PRobe far-Infrared Mission for Astrophysics) is one
of two probe concepts currently studied by NASA, which could be operational in
the 2030ies. Even assuming that PRIMA flies, its lifetime is limited to
5 years, too short to put meaningful limits on the lifetimes of decade-long
FUOr-type outbursts.

Only ground-based facilities can provide decade-long
monitoring. While demanding, observing at wavelengths as close as possible
to the far-IR regime is motivated by the expected stronger amplitude of the
flux variation with accretion rate change at shorter sub-millimeter wavelenghts
and the better contrast between the warmer protostars and the colder
surrounding dust. Observations at
350~$\mu$m (ALMA Band~10) will be possible for a significant
fraction of the total available observing time at the best high altitude, dry
sites such as Chajnantor plateau, allowing for repeated, wide field, sensitive
observations of large samples at 350~$\mu$m.

\noindent{\bf Why a single dish telescope is needed:} ALMA (or any other
interferometer)
is not well suited for tracing sub-millimeter variability. In particular
for large samples (i.e., short integration times, poor uv coverage),
varying configurations and observing at varying hour-angle
imply varying spatial filtering, making it impossible to unambigously
attribute measured flux variations to real brightness variations of the
protostars. In addition, the surrounding emission is used as flux
reference to substantially improve the accuracy and sensitivity to variability
\cite{Mairs2024},
which would also be compromised by varying spatial filtering by an
interferometer.

\noindent{\bf Why a large field-of-view is needed:} presently protostar
variability surveys covered areas of a few square degrees containing a few
hundred protostars in the most nearby regions and a few selected more massive
and distant regions. \cite{Lee2025} argue, that sample sizes on the order of
2000 (low mass) objects are needed; this would yield on the order of a few
ten outburst detections and meaningfully low uncertainties to constrain the
frequency of
outbursts (within the 5~yr duration of PRIMA).

\noindent{\bf Why decent spatial resolution is needed:} in particular for
(typically distant) clustered regions containing high mass protostars
spatial resolution increases the contrast between the variable protostar
and the non-variable surrounding protostars and background clump emission.
It also provides for more accurate cross-identification with higher resolution
(e.g., ALMA) observations.


\section{Technical requirements}


The described science goal requires decades long monitoring of thousands of
protostellar sources in nearby star forming regions and the Galactic plane.
Observing as close as possible (from the ground) to the far-IR peak of
protostellar SEDs around 100~$\mu$m requires a push towards the
short-wavelength/high frequency
part of the sub-millimeter regime while good spatial resolution (few arcseconds)
helps to increase the contrast against the surrounding emission and to pinpoint
the variable sources in clustered regions. Efficient mapping requires a
combination of large field-of-view and instantaneous sensitivity.



The AtLAST concept \cite{Mroczkowski2025} for a 50~m diameter single dish
with an instantaneous
FOV of up to 2$^{\circ}$ diameter fullfills these requirements. Located near ALMA
at a 5000~m altitude site it will operate up to a frequency of 950~GHz
($\sim$300~$\mu$m, i.e., ALMA Band~10) and will provide a diffraction limited resolution of
$\sim$1.$\!^{\prime\prime}$5 at the shortest wavelenghts. The large dish combined
with the large FoV provides for fast mapping: the mapping speed will be up
to 10$^5$ times better than ALMA, while AtLAST's continuum sensitivity is
comparable to that expected from the entire ALMA array post-WSU.


AtLAST's combination of a large FoV and a large aperture, i.e., high
instantaneous sensitivity, will revolutionarize sub-millimeter transient
science \cite{Orlo2025}. Particularly the prospect of efficient observations
in the high-frequency sub-millimeter regime, up to 350~$\mu$m/ALMA Band~10, is
unprecedented. AtLAST will thus be uniquely positioned to decipher protostellar
accretion and its temporal variability.


\vspace*{2mm}

\noindent {\bf References:} \printbibliography[heading=none]

\end{document}